\documentclass [A4; 9,5pt]{article}
\usepackage {graphicx}
\usepackage {longtable}
\usepackage[left=3cm,right=3cm,
top=2cm,bottom=2cm,bindingoffset=0cm]{geometry}

\begin{document}

\begin{center}
\textbf{Journal of Physical Science and Application}
\end{center}

\vspace{1cm}

\begin{center}
\textbf{\Large{On High Brightness Temperature of Pulsar Giant Pulses}}
\end{center}

\vspace{1cm}

\begin{center}
Victor M. Kontorovich
\end{center}

\vspace{0,3cm}

\begin{center}
\textit{Decametric wave department, Institute of Radio Astronomy NAS of 
Ukraine, Kharkov 61002, Ukraine;}
\textit{Radio Physics, Mechanics and Mathematics departments, V.N.Karazin 
National University, Kharkov 61022, Ukraine}
\end{center}

\vspace{1cm}

Received: August 02, 2014 / Accepted: September 06, 2014 / Published: 
November 10, 2014.

\vspace{1cm}

\textbf{Abstract}: A wide range of events observed at the giant pulses (high 
energy density, observed localization of GPs relative to the average pulse, 
fine structure of GPs with duration up to some nanoseconds, observed 
circular polarization of GPs, correlation between the GP phase and the phase 
of the hard pulsar emission - X-ray and gamma) can be explained from the 
viewpoint that the internal polar gap is a cavity-resonator stimulated by 
discharges and radiating through the breaks in the magnetosphere. The new 
results in this field (the em waves generation in the gap in the process of 
longitudinal acceleration in the electric field vanishing on the star 
surface, high frequency break in the spectrum as a result of switching off 
this generation, formation in this process a power-low spectrum with a hf 
break, the possibility determination of pulsar magnetic field by the hf 
break position, the difference between main pulse and inter pulse mechanism 
generation, quantization of em tornado rotation in the gap and appearance of 
the bands in the inter pulse spectrum, influence the high energy density in 
the gap on pair generation and position of the dead line in pulsars) are 
added in the Intermediate\textbf{} Epilogue.

\vspace{1cm}



\vspace{1cm}


\noindent\textbf{Nomenclature}

\vspace{0,5cm}
\noindent
GP: giant pulse

\noindent
hf: high frequency

\noindent
em: electromagnetic

\vspace{0,5cm}
\noindent\textbf{Greek letters}

\vspace{0,5cm}

\_\_\_\_\_\_\_\_\_\_\_\_\_\_\_\_\_\_\_\_\_\_\_\_\_\_

\textbf{Corresponding author:} \noindent Victor M. Kontorovich,\\
\noindent doctor of sci, professor, research fields: theoretical\\
physics, plasma physics, astrophysics. \\
\noindent E-mail: vkont1001@yahoo.com, vkont@rian.kharkov.ua

\newpage

{\large{Introduction}}

\bigskip

Giant pulses (GP), sporadically observed in a small number of
pulsars\footnote{ Crab (B0531+21), B1937+21, B1821-24, B0540-69,
B1112+50, B1957+20, J0218+4232, J1823-3021A, B0031-07, J1752+2359,
B0656+14$^{ 6, 31, 33- 35, 37-42} $}$^)$, are a riddle which has
not yet been solved  (see reviews$^{1-5}$). GP is characterized by
enormous flux density$^{6}$, extremely small pulse duration (down
to a few nanoseconds)$^{7}$, presence of circular polarization of
both directions$^{8}$, power distribution by energies$^{9}$,
mainly location in the narrow window with respect to the average
pulse position$^{10}$ and correlation between the GP phase and the
phase of the hard pulsar emission (X-ray and $\gamma
$-ray)$^{11-12}$. All these features fundamentally distinguish GPs
from ordinary pulses\footnote{ See also the works
$^{43-48}$.}$^)$. Nevertheless, GP seems to be "a frequent, but
rarely observed phenomenon inherent in all pulsars"$^{10}$. A
number of pulsars\footnote{ B0809+74, B0823+26, B0834+06,
B0943+10, B0950+08, B1133+16}$^)$ emits anomalously intensive
pulses$^{13}$ which by their properties seemingly do not differ
from GPs. Trying to explain GPs by plasma mechanisms in
magnetosphere in which different variants of two-stream
instability are realized$^{14}$ needs considering strongly
nonlinear effects such as modulation instability$^{15-16}$,
Zakharov plasma wave collapse (the more popular!)$^{8}$,
reconnection of the magnetic field lines$^{17,18}$, induced
scattering in narrow beams$^{19}$, etc. (see references in the
reviews$^{5}$).

When explaining the GPs in these works, we have seen that the
pulsar is regarded as a "plasmic generator", a "device", in which
the radiation processes are governed by different (nonlinear)
processes in the magnetosphere plasma.

In work $^{10}$ it is indicated that GPs are characterized by the
extremely high energy density of order 10$^{15}$ erg/cm$^{3}$,
which appears as a key moment and forms the basis of this study.

It should be emphasized, and this is the main subject-matter of
this work, that in the similar terms the pulsar can be thought of
as a "vacuum device", in which the magnetospheric plasma acts as
the walls, limiting the cavity and the waveguides as well the
breaks in the magnetosphere, through which the radiation of GPs is
transmitted. The radiation itself arises at discharges in the
inner gap.

Thus we proceed from the idea that the pulsar internal polar
gap$^{20}$, where the particle acceleration processes occur in the
longitudinal electric field$^{21}$, is a cavity-resonator (with
respect to the low-frequency radiation$^{22,23}$). In the
work$^{22}$ the idea of resonator was illustrated by analogy with
the resonator "earth-ionosphere" excited by the lightning
discharges: "We live in a resonant cavity bounded below by the
earth and above by the ionosphere. Lightning strikes excite
low-frequency electromagnetic Schumann resonances in this cavity,
as predicted by Schumann (1952), with a fundamental frequency of
about 8 Hz. Such behaviour should also occur above the magnetic
poles of pulsars due to the well-defined boundaries\footnote{ See
also discussion in Appendix A}$^)$ found there"$^{22}$. The modes
of the cylindrical resonator were used in$^{22}$ to describe such
feature as a "carusel" structure of the pulsar micro pulses, and,
as seen from the quotation above, the analogy with Schuman
resonance in the spherical "earth-ionosphere" resonator (see e.g.
the book$^{24}$ of P.Bliokh and his collaborators) was discussed.

In the work $^{23}$ another aspect of the problem is brought to
the foreground: the stationary random process of discharges on the
polar cap surface and, as a result, formation of a stable average
shape of the pulse. That was demonstrated with the aid of the toy
dice model. The idea of the resonator and waveguides was also
formulated and then used in the context of pulsar radio and hard
radiation$^{25}$ from the polar gap. Owing to this idea the
radiation emanates from the gap through the waveguides in the
neighborhood of the magnetic axis and through the slots$^{26}$ at
the open field line borders. The radiation also leaks out through
the magnetosphere plasma. The correlation between radio and gamma
radiation, arising due to inverse Compton scattering of
accelerated electrons at a powerful low-frequency radiation, can
be an indirect confirmation of the powerful oscillations in the
gap$^{25}$. GP may be another and more evident manifestation of
the strong oscillations in the gap.

In this work we distinguish between two approaches, that we
propose. One of them is to consider the passage of powerful
quasi-stationary oscillations in the cavity through the random
breaks in the magnetosphere. Under the assumption, the inner gap
serves as a resonator (see Section 2 and Appendix B). With the
other approach we examine (to explain the GP microstructure) the
direct radiation of the individual discharge by-passing  the
plasma through the slot and the waveguide. This allows explain
both the nanosecond pulse duration (due to relativistic aberration
in the fast-rotating pulsars) and the observed circular
polarization of GPs (see Section 3 and Appendix C). The last
approach does not use the concept of the polar gap as a cavity.
Similar notions and views may turn out to be useful in analyzing
the properties of ordinary pulses, their substructure, subpulse
drift, etc. However, these attempts will doubtless be restricted
to the lack of developed concepts of the "random" magnetosphere of
open field lines and the most probable routes of the radiation
transmission through such  a medium. These questions are briefly
discussed in Appendix A in the application to the random
magnetosphere which is a stationary one on average only.

\bigskip

{\large {Description of the model}}

\bigskip

In our opinion, GP can represent a direct emanation of radiation
from the gap (fig.1) through the breaks in the magnetosphere.

The particles accelerated in the gap undergo Compton losses, which
exceed the curvature radiation losses under sufficient density of
oscillation energy$^{25}$. The energy received from the
longitudinal electric field and, in the long run, from star
rotation, is emanated in the form of gamma quanta (which generate
the electron-positron plasma), and in the form of radio emission
as in common models$^{27-28}$.

\begin{figure}
 \includegraphics[angle=0,scale=0.5]{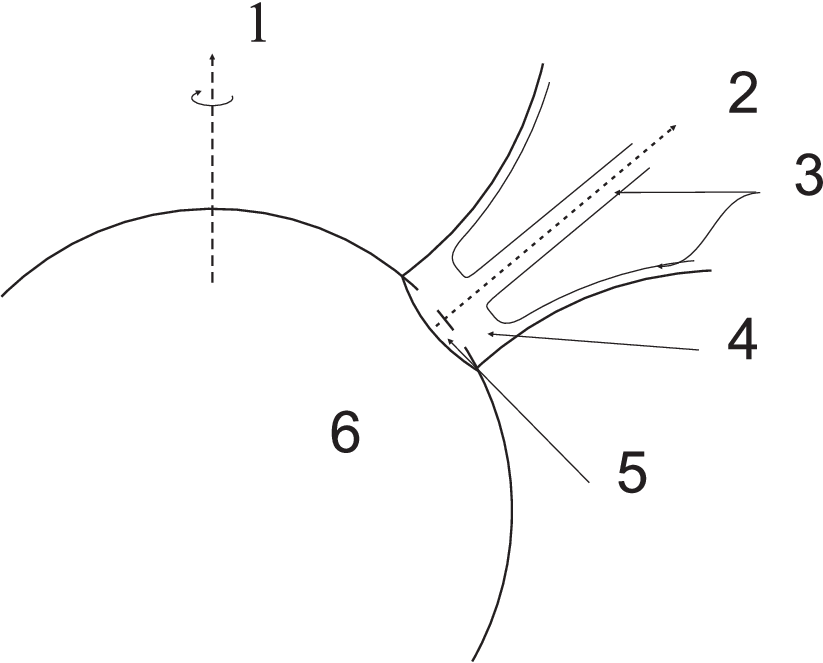}
 \caption{The inner polar gap serves as a cavity-resonator for
radio band electromagnetic oscillations excited by discharges in
the gap. The radiation goes out through the wave-guides and
percolates through the magnetosphere plasma. 1 -- rotation axis, 2
-- magnetic axis, 3 -- waveguides, 4 -- polar gap, 5 -- polar
cap, 6 -- neutron star.}\label{fig1}
\end{figure}

The energy density $U$ in the gap can be estimated according to
the law of electromagnetic field energy conservation in the cavity
(see Appendix B) stimulated by external discharge currents in view
of the radiation losses (the irradiation through the waveguide).
Under the stationary conditions the square averages of these terms
(the averages over space and time) have to be equal to each other.
To estimate the scalar product of the current density and the
electric field we use the Cauchy inequality. The energy density is
determined as the square of the average electric field. We find
the upper limit of energy density from the condition that the
average current through the gap should be equal to the
Goldreich-Julian current, which results in the following$^{25}$

\begin{equation}
\label{eq1}
U \le 4\pi \left( {1 + \mu}  \right)^{2}\left( {\Sigma _{PC} \;/\Sigma _{w}
} \right)^{2}\left( {h \cdot \rho _{GJ}}  \right)^{2},
\end{equation}

\noindent where $\Sigma _{PC} $ is the area of the polar cap,
$\Sigma _{W} $  is the area of the waveguide cross-section, $\rho
_{GJ} $ is the density charge of Goldreich-Julian, \textit{h} is
the average height of the gap, $\mu $ is the part of the resonator
locked modes that cannot go out through the waveguides. The
appropriate parameter estimation leads to the energy densities
comparable with those observed at GPs: $U \le 10^{16}erg/c$. The
same condition may be transformed into the energy density
restriction from below if we express it in the terms of radiation
intensity $I_{R} $:

\begin{equation}
\label{eq2}
8\pi U \ge \left( {I_{R} /c\rho _{GJ} h \cdot \Sigma _{pc}}  \right)^{2},
\end{equation}

\noindent which allows one to estimate $U \ge 10^{12}erg/s$. One
more estimate of the energy density \textit{U} in the gap may be
made by using the momentum conservation law. The very high level
of the energy density in the gap means that the pressure of
radiation in it has to be very high too. It follows that the form
of the boundary of the cavity with plasma must be sufficiently
sharp. The boundary condition of the momentum density flux
continuity gives \footnote{ Cf. with the non-relativistic
expression for momentum density flux $P + mnV^{2}$, \textit{P} is
the pressure, \textit{n} is the particle concentration, \textit{m}
is the mass of the particle.}$^)$ for \textit{U}:

\begin{equation}
\label{eq3}
U \le mn_{GJ} c^{2}\Gamma _{sec}^{2}
\end{equation}

\noindent where $n_{GJ} = \rho _{GJ} /e$ is the Goldreich-Julian
concentration, $\kappa \approx 10^{3}$  the multiplicity factor,
$\Gamma _{sec} $  the Lorentz-factor of the secondary $e^{ \pm}
$-plasma above the inner gap. Here we assume that the
contribution of gamma-ray quanta in the flux is  of the same order
or smaller. It implies that the position of the control section in
the plasma must be at a distance of some gamma quanta mean free
pass above the boundary. The estimation $\Gamma _{sec} \approx
10^{3} - 10^{4}$ gives us $U \le 10^{14} - 10^{15}$ erg/cm$^3$ --
the same as the previous ones on the rough scale. The more
accurate estimation requires the solution of difficult problems of
the plasma boundary structure and of the transmission (reflection)
of the powerful electromagnetic waves through such a boundary.
This problem is  not yet solved. (In that case the linear approach
to the problem is not applicable at all.)

Thus, we assume that a great amount of energy may actually be
concentrated in the polar gap and the gap may be regarded as a
cavity resonator. Some additional
arguments in favor of this will be mentioned in the text below. We
will discuss below the implications of these assumptions relative
to the GPs and will confirm a consistency (noncontradictory) and
efficiency of such a hypothesis.

The energy radiated by the GPs is determined by the oscillation
energy density in the gap and parameters of the break in the
magnetosphere plasma. The energy emanated in the pulse duration is
proportional to the volume of the break $S \cdot \Delta z$, where
$\Delta z$ is its height proportional to radiation time $\Delta
t$.

\bigskip

{\large{Discussion}}

\bigskip

For the distribution of the oscillation energy density in
frequencies $U\left( {\nu}  \right)$ we take the same power law
with index $\alpha _{R} > 1$ as for the observed radio frequency
radiation\footnote{ Pulsar in the given model emanates GPs in the
frequency interval $\nu _{1} < \nu < \nu _{2} $, where $\nu _{1}
\approx c/\sqrt {S} $ is determined by the conditions of the wave
transmission through the break (waveguide) of the given
cross-section \textit{S}, $\nu _{min} < \nu < \nu _{1} $ is the
interval of the locked modes.}$^)$

\begin{equation}
\label{eq4}
U\left( {\nu}  \right) = U \cdot {\textstyle{{\left( {\alpha _{R} - 1}
\right)} \over {\nu _{min}} }}\left( {{\textstyle{{\nu _{min}}  \over {\nu
}}}} \right)^{\alpha _{R}} ,
\quad
U = \int_{\nu _{min}} ^{\nu _{2}}  {d\nu \,U\left( {\nu}  \right)} .
\end{equation}

Here $\nu _{min} \approx c/\lambda _{max} $ is determined by the
longest-wave modes, and $\nu _{2} $ is the maximal possible
frequency of oscillations in the cavity corresponding to the
high-frequency edge of the pulsar radio spectrum. The link between
the break area \textit{S} and received flux density $F\left( {\nu}
\right)$ at  the given frequency has the form

\begin{equation}
\label{eq5}
U\left( {\nu}  \right) \cdot c \cdot S = D^{2} \cdot \Delta \Omega \cdot
F\left( {\nu}  \right).
\end{equation}

Here \textit{D} is the distance from the pulsar, $\Delta \Omega $
is the spatial angle interval in which the radiation is emanated.
The azimuth angular distance $\Delta \varphi $ is passed for the
time $\Delta t = P\Delta \varphi /2\pi $ where \textit{P} is the
period. Thus, the solid angle $\Delta \Omega = \Delta \theta \cdot
\Delta \varphi $ correlates with the time of radiation $\Delta t$
as $\Delta \Omega \approx 2\pi \cdot \Delta \theta \cdot \Delta
t/P$. The break area $S$ also requires the time $\Delta t$ for
passing through. Taking into consideration $S < < \pi R_{s}^{2} $,
$S = \pi R_{s}^{2} \Delta \Omega _{s} $ where $R_{s} $ is the
effective radius  where the break is realized and $\Delta \Omega
_{s} $ is the corresponding spatial angle, we see that the length
of the pulse $\Delta t$ falls out from our relations. If $\Delta
\theta _{s} \approx \Delta \theta $ then we have $\pi R_{s}^{2} =
D^{2} \cdot F\left( {\nu} \right)/\left( {U\left( {\nu} \right)
\cdot c} \right)$ or

\begin{equation}
\label{eq6}
\pi R_{s}^{2} = \frac{{D^{2} \cdot F\left( {\nu}  \right) \cdot \nu _{min}
}}{{\left( {\alpha _{R} - 1} \right) \cdot U \cdot c}} \cdot \left(
{\frac{{\nu} }{{\nu _{min}} }} \right)^{\alpha _{R}} .
\end{equation}

\noindent The average GP profile is attached to the definite phase
of the polar cap $\theta $. Hence, the average shape of the arch
of the cavity is $h\left( {\theta} \right) \approx R_{S} -
R_{\ast} $. Probably, close to the magnetic axis this description
is rather acceptable. Yet, near the slot it is necessary to take
account of its curvature, a decrease in amplitude of the low
frequency oscillation near the border of the open field lines,
along with other complicating factors$^{30}$.

Now we can make the independent rough estimate $^{29}$ of the
energy density \textit{U} (from the top). Assume (only for this
estimation!) that it makes the same contribution to each spectral
interval of the observable density flux. If we make a sufficiently
rough model to be considered in each frequency window $U = Const$,
then for the spectral index $\alpha = 3$ the area of the break
falls out (see $^{29}$) from formulas $F\left( {\nu} \right) =
\left( {\alpha - 1} \right) \cdot S^{\frac{{3 - \alpha} }{{2}}}
\cdot U \cdot \lambda ^{\alpha} /\left( {D^{2} \cdot \Delta
\Omega} \right)$. Here \textit{D} is the distance to the pulsar,
$F\left( {\nu}  \right)$ is the flux at a frequency $\nu $,
$\lambda $ is the wavelength, $\Delta \Omega $ is the solid angle
of radiation. Since $\alpha \approx 3$ corresponds to the spectrum
of the Crab pulsar, we are able to evaluate $U$ from the
observational data for Crab: $U = D^{2}\Delta \Omega \cdot \lambda
^{ - 3}F\left( {\nu} \right)/2\,\,\,,\,\,\,\alpha = 3$ (for this
estimation only!). This gives us $U \le 10^{17}erg/cm^{3}$, which
is in good correlation with other estimates (see the text above
and App.B).

\bigskip

\textit{GP short duration}

\medskip

An extremely short giant pulse duration of several nanoseconds
allows us to suggest that the pulses arise in the inner
gap$^{12}$  in the process of primary electron acceleration
$^{13}$ to the gamma-factors of order of 10$^{7}$. Indeed,
relativistic aberration (see also the work $^{49}$) in the primary
electron beam reduces the cone of radiation down up to the values
$\delta \varphi \sim 10^{ - 7}$, and in this case the rotation
with periods $P \approx 2\pi \cdot 10^{ - 2}$s leads to the
nanosecond pulse duration $\delta t \approx \delta \varphi \cdot
P/2\pi \sim 10^{ - 9}$s. This explanation is also consistent with
the fact that giant pulses are observed  in the rapidly rotating
pulsars only. Thus, we have to deal with the emission of an
individual discharge in the inner gap (cf. $^{28}$).

\bigskip

\textit{GP phase}

\medskip

The observed localization of the GP phase can be associated with
radiation through the waveguides. In pulsar B1112+50 GPs are
located in the center of the average pulse$^{31}$. We reckon that
this localization may correspond to the radiation through the
"waveguide" near the magnetic axis of the pulsar. If the GP phase
corresponds to the "edge" of the average pulse, then it most
likely corresponds to the radiation through the slot. The edge can
be either retarded against the average profile$^{9,32}$
(B1937+21), or advanced$^{33}$ (J1823-3021A). This corresponds to
trailing or leading edges of the slots in the section of the
telescope diagram. The fine structure of the GPs may reflect the
discreteness of the discharges$^{28}$  visible through the breaks.

From this point of view the correlation between the GP phase and
the phase of the hard pulsar emission (X-ray and $\gamma $-ray)
$^{1,11,12}$ becomes  truly evident:  radiation arises in the same
process of particle acceleration and goes out through the same
waveguides.

Localization of GPs (near the waveguides and near the slots)
signifies that the magnetosphere of the open field lines is not
transparent, on average, to radiation, except for these places of
GP localization. This statement also supports the idea that the
inner polar gap is a cavity resonator. However, sporadic GP can
occur at all phases$^{45}$ .

\bigskip

\textit{Circular polarization and electromagnetic tornado}

\medskip

The observed circular polarization$^{8, 32} $  is naturally
explained by the peculiarities of the discharge in the inner gap.
Coulomb charge repulsion in the discharge bunch furnishes the
radial electric field orthogonal to the magnetic one. Owing to the
drift in the crossed fields it causes the discharge jet rotation
around its axis (see Appendix C) and, accordingly, the circular
polarization of generated electromagnetic waves. In fact, both
accidental and regular deviations from the ideal axisymmetric form
of discharges results in the rotational electric field around the
discharge axes and the related circular polarization of radiation.
Owing to the drift the discharge channel turns into a peculiar
vortex resembling the well-known tornado. However, contrary to the
hydrodynamic nature of the usual tornado, the inner gap tornado
is of purely electromagnetic origin.

\medskip

\begin{figure}
 \includegraphics[angle=0,scale=0.5]{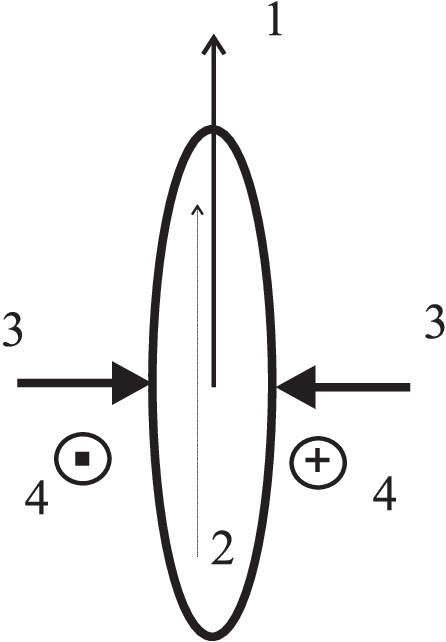}
 \caption{In the discharge bunch filament the Coulomb particle
repulsion leads to rotation in the crossed electric and magnetic
fields around the bunch axis. This rotation transforms the bunch
into an electromagnetic tornado. For electrons and positrons the
rotation directions are opposite. The rotation can cause the
circular polarization which is observed in GPs. 1 -- magnetic
field, 2 -- bunch velocity, 3 -- radial electric field of space
charge, 4 -- drift velocities. }\label{fig2}
\end{figure}

\medskip

From the equation $divE = 4\pi \rho $ for the spatial charge
$\rho$ of the form

\begin{equation}
\label{eq7}
\rho \left( {r} \right) = \rho _{0} r_{0}^{4} /\left( {r_{0}^{4} + r^{4}}
\right),
\end{equation}

\noindent the Coulomb field $E_{r} $ radial to the vortex axis  is

\begin{equation}
\label{eq8} rE_{r} = 2\pi \rho _{0} r_{0}^{2} arctg\left( {r^{2}
/r_{0}^{2}} \right).
\end{equation}

For the particle drift velocity (fig.2)

\begin{equation}
\label{eq9}
V_{\varphi}  = cE_{r} /H
\end{equation}

\noindent at large distances from the axis there is a movement
with constant circulation and solid-type rotation on the small
ones.

The quasi-classical quantization is possible by using the condition

\begin{equation}
\label{eq10}
mrV_{\varphi}  = n \cdot \hbar ,
\end{equation}

\noindent
which leads to the angular frequency of the bunch

\begin{equation}
\label{eq11}
\Omega _{n} = n \cdot \hbar /mr^{2}.
\end{equation}

The particle linear density $N = \pi r_{0}^{2} \cdot \rho _{0} /e$
and, correspondingly, the current of the vortex tube \textit{ceN}
are quantized. As a result,  they do  not contain any model
parameters and are dependent on the magnetic field only:

\begin{equation}
\label{eq12}
ceN = n \cdot \hbar B/m.
\end{equation}

Here \textit{n} is integer or half-integer. The full
Goldreigh-Julian current flowing through the gap is provided by
\textit{q} vortex lines as

\begin{equation}
\label{eq13}
q \simeq \frac{{\Omega \,m}}{{\pi n \cdot \hbar} }\Sigma _{PC} \quad .
\end{equation}

In the ground state it is required  that $q \approx 10^{7}$ lines.
For a smaller \textit{q} the Rydberg states with $n > > 1$ are
required. Discharges may arise owing to the charges running off
from the microscopic edges of the polar cap surface, thereby
forming bunches and low frequency radiation that feeds the cavity.
The edges may be destroyed in this process followed by subsequent
restoration. Rotation frequencies form the bands whose borders are
determined by tornado internal and external radii. For example,
the radius $r \approx 10^{ - 5}$cm matches to the frequency of
$\Omega \approx 10^{10}$s$^{-1}$. Such a structure might explain
the frequency bands$^{34}$  observed in the GP spectrum.\footnote{
These strips have been observed in the interpulse spectrum only.
According to $^{18}$
the strips correspond to Bernstine modes.}$^)$

\bigskip

\textit{Power-law distributions}

\medskip

In the scheme under consideration the energy emanated in the GP is
proportional to the break area. The observed power-law of GPs
occurrence frequency  in accordance with their energy$^{35}$ means
the power dependence of a break appearance probability upon the
break area. Here the analogy with the geophysical phenomena in
which many examples of such distributions are known seems to be
quite relevant. Many observed statistical patterns causing
power-low distributions can be obtained under the elementary
assumption of the small correlation time of random force with
different physical definition of "forces" and "particles"$^{36}$.
The consequence of such a force correlation is the dependence of
the form $ N\left( { \ge E} \right) \propto E^{ - 1}$ for the
cumulative frequency of the events with "energy" \textit{E} which
explains many empirical dependencies$^{36}$. Here, apparently, it
is necessary to add frequency of the GP pulsar appearance where
the energy is proportional to the break areas in the
magnetosphere. With smaller energies \textit{E} the GP complies
with the power-law$^{36}$  $\quad N\left( { \ge E} \right) \propto
E^{ - a}$ (\textit{a} =\textit{0.9 -- 1.1}). With greater energies
a sharp bend in the distribution is observed. It can easily be
interpreted as the result of the break superposition which leads
to the steepening of the distribution ($a>2$).

\bigskip

{\large{Resume}}

\medskip

Now we summarize the foregoing.

In the giant pulses the energy density is comparable with that of
oscillation in the inner gap. It means that we
obtain radiation at the GP instants  directly from the gap
by-passing the magnetosphere plasma. It becomes possible if the
breaks or the cracks and holes appear in magnetosphere due to
random allocation of plasma-generating discharges. GP duration is
not always fully attributable to its physical nature. The pulse
length may be determined by the dynamics of the break appearance
and disappearance in the magnetosphere. The GP power-law intensity
distribution is also determined by the probability of appearing
the breaks of different size and is not directly associated with
GP radiation physics. The appearance of breaks is most probable in
the phases that conform to the "waveguides" near the magnetic axis
or near the edge of open field lines. Separate discharges may be
superimposed on the radiation through the breaks. In such cases
the relativistic aberration (see also $^{49}$) narrows down the
angle range $\Delta \varphi $ up to 10$^{-7}$ and fast rotation
contracts the pulse duration $\Delta t = P\Delta \varphi /2\pi $
up to 10$^{-9}$s. Obviously, circular polarization of GP is also
governed by physics of its radiation. Such a polarization can be
generated by electromagnetic tornados in the gap.

To sum up consider a simple analogy with the Sun radiation
observed on a cloudy day. In the breaks in the clouds we observe
some kind of a "pulse". A large break lets more energy get through
(an analogue of the GP intensity) but it is the characteristic of
the break and not of the emitter. The "pulse" duration is
determined by the break existence but not by the physics of the
radiation. The steepness of the "pulse" is also determined by the
movement of the clouds but not by the solar processes. The
emitter's main characteristic is the brightness temperature and
spectrum as well as polarization. But fast fluctuations such as
solar flares and bursts of solar radiation could make their own
contribution if we can see them between the clouds.

\medskip

{\large{Acknowledgements}}

\medskip

I am grateful to V.V.Usov for support of the idea of this work, to
V.S.Beskin and Ya.N.Istomin for calling my attention to the paper
$^{10}$, and to O.M.Ul'yanov for helpful discussion and comments
on the text.

\bigskip
{\large{Appendix A. On nontransparency of the pulsar magnetosphere
in the open field lines region}}
\medskip

Electrodynamics of the boundary of the polar gap with plasma in
the region of open field lines, which limits the gap from
above, is a fairly challenging task and has not been sufficiently
developed, although it was announced in the work$^{22}$. Therefore
we give below the following physical considerations indicating
that this boundary should have relatively high reflective
properties, so that the gap could be considered as a good
resonator. Now we would like to focus our attention primarily on
the random environment (the electron-positron plasma) which has to
pass through the wave from the inner gap before leaving it.

This randomness is the result of random character of the
discharges on the surface of the polar cap$^{23}$. Discharges
generate the plasma via the birth of electron-positron pairs in a
magnetic field (by gamma-ray photons emitted at breakdowns). The
geometry of the open field lines and the strong magnetic field
create a configuration that is close to the one-dimensional one,
in which the predominant role is played by the backward scattering
that results in Anderson localization$^{50}$. With such scattering
in a stationary random environment the reflected wave travels the
path along which all the random phase shifts acquired by the wave
along the straight path are fully compensated for. This leads to
an exponentially small average transmission coefficient of the
wave through a layer of random medium exceeding the localization
length$^{51}$ $l$  and, accordingly, to an exponentially small
difference between the average reflection coefficient $\overline
{R} $ and unity$^{52,53}$ :

\[
\overline {R} \approx 1 - \frac{{\pi ^{5/2}}}{{2}}\left( {\frac{{L}}{{l}}}
\right)^{ - 3/2}exp\left( { - \frac{{L}}{{4l}}} \right),
\]

\noindent where $L$ is the thickness of the layer. (The wave
transmission occurs only on the most favorable paths that are
different for different random realizations$^{54}$; they are
reported in the text in a  simplified manner as the breaks in the
magnetosphere). In order to render the use of these findings
reliable it is necessary to make sure that the situation is
actually quite close to one-dimensional and that the
nonstationarity of configurations (in the unchallenged stationary
random process) does not destroy the desired coherence of
scattering. (Nonstationarity may occur even on average$^{43}$).
This question is left open now and we hope to return to this issue
separately.

This does not contradict the fact that in a smoothly inhomogeneous
magnetized relativistic electron-positron plasma with identical
distributions of  electrons and positrons the transverse
electromagnetic waves have the gapless nature$^{55-57}$ and,
seemingly, can easily penetrate through the
magnetosphere\footnote{ The most frequent objection to the
resonance properties of the inner gap. Let us note, that the
waves are linearly polarized.}$^)$. However, the magnetosphere is
not the smoothly inhomogeneous, especially in the area of open
field lines, and this is a basic fact for the problem of
reflection of electromagnetic waves. But besides this, in the
transition layer between the gap and the plasma the
electromagnetic waves is of a different character. Indeed, there
is a strong longitudinal electric field in the transition layer,
which turns the positrons, that are born in pairs, back to the
star.

Therefore, there should be a layer of non-relativistic positron
plasma with the Goldreich-Julian plasma frequency $\omega _{GJ} =
\sqrt {\omega _{c} \Omega}  \approx 10^{10}s^{ - 1}$ (here $\omega
_{c} $ is the cyclotron frequency and $\Omega $ is the angular
frequency of star rotation). Note that the upper edge of the
pulsar radio spectrum, as remarked in $^{25}$, is located close to
this frequency. It is important, however, to highlight one more
circumstance. The back scattering is strongly suppressed in the
relativistic outflowing plasma. But in the transition layer in
which the reverse flow of the positrons is accelerated up to the
speed of light, the backscattering of the waves is also very
efficient due to the same relativistic aberration.  This should
lead to the opacity of the plasma and to the high reflective
properties of the upper lid of the gap resonator. It will
be recalled that the existence of the cavity is one of
assumptions in this paper. We analyze the consequences of this
assumption, which in fact allows us to give a consistent
explanation of all the facts relating to the issue of GPs.

\bigskip
{\large{Appendix B. Energy density estimation}}
\medskip

The average energy density \textit{U} of the low-frequency
electromagnetic field\textit{} in the gap can be estimated from
the energy conservation law in the cavity excited by external
discharge currents. The conservation law averaged over time yields
the balance condition $div\bf {S} = - \bf {j_{ex} E_{\sim}} $
where $\bf {S} $ is the Poynting vector, $\bf {j_{ex}} $ is the
density of the spark currents and $\bf {E_{\sim}} $ is the
strength of the low-frequency electric field. The bar denotes
averaging over time. By the average values of $\overline {A} $ we
imply their mean square values $\sqrt {\overline {A^{2}}} $.
Integrating the balance equation over the volume of the resonator
and using the Gauss' theorem we obtain $\oint \overline {\bf S} d
\bf {\Sigma} = - \int$ $dV \overline {\bf {j_{ex} E_{\sim}} } $.
The average value $\,\int dV \overline { \bf { {j_{ex} E_{\sim}}}}
$ is estimated with Cauchy-Bunyakovsky inequality $\,\int {dV}
\overline { \bf {{j_{ex} E_{\sim}}}}  \le \int {dV} \overline
{j_{ex}} \cdot \overline {E_{\sim}}  $. For the average current
density $\overline {j_{ex}} $ we take the Goldreich-Julian value
$\overline {j_{ex}}  = c\rho _{GJ} $ where the Goldreich-Julian
charge density is $\rho _{GJ} = - \bf{\Omega B}$/$\left( {2\pi c}
\right)$. Using the theorem about the average for the integral in
the right-hand side of the balance condition inequality we obtain
the estimate $\int {\overline{\bf {j_{ex} E_{\sim}}} }  dV \le
c\rho _{GJ} \cdot \overline {E_{\sim}} \cdot \Sigma _{PC} h$. Here
\textit{h }is the height of the gap, $\Sigma _{PC} $ is the area
of the polar cap. Note that the power emitted in the
radio-frequency region is $I_{R} = \oint
\bf{{\overline{S}d\Sigma}}$ = $cU_{R} \Sigma_{w} $, where $U_{R} =
U/\left( {1 + \mu} \right), \quad U=\overline {E_{\sim}^2} /\left(
{4\pi} \right)$ and the contribution of the closed modes is $\mu
\approx \;\left(\omega _1/\omega _{min} \right)^{\alpha _{R} -
1}\left[ {1 - \left( \omega _{min}/\omega _1 \right)^{\alpha _{R}
- 1}} \right]$. Inserting these relations into the balance
condition we have $\frac{{1}}{{1 + \mu} }c\frac{{\overline
{E_{\sim}^{2}} } }{{4\pi} }\Sigma _{w} \le c\rho _{GJ} \overline
{E_{\sim}}  \Sigma _{PC} h$ and whence we estimate the average
electric field in the gap: $\overline {E_{\sim}}  \le 4\pi \rho
_{GJ} h\frac{{\Sigma _{PC}} }{{\Sigma _{w}} }\left( {1 + \mu}
\right)$. Finally, raising to the square we find the upper bound
for the energy density in the gap (1)
$$U = \frac{{\overline {E_{\sim}^{2}} }}{{4\pi} } \le 4\pi
\left( {1 + \mu} \right)^{2}\left( {\Sigma _{PC} /\Sigma _{w}}
\right)^{2}\left( {h\rho _{GJ} \,} \right)^{2},$$ whence follows
under the typical values of parameters that
$$U \le 10^{16}\left({1 + \mu} \right)^{2}\;erg/cm^{3}.$$
On the other hand, we can make an independent estimate of
\textit{U} by expressing the average field $\overline {E_{\sim}} $
through the intensity of radio emission $I_{R} $. Using the
balance condition in the form $\oint
\bf{\overline{S}}d\bf{\Sigma}$ $\le c\rho_{GJ} \cdot \overline
{E_{\sim}} \cdot \Sigma_{PC} h$ we have  estimate $\overline
{E_{\sim}}  $ limited from below $I_{R} \le c\rho _{GJ} \Sigma
_{PC} h \cdot \overline {E_{\sim}} \,$which leads  to the lower
bound for the energy density of oscillations in the gap (2):
$$\frac{{1}}{{4\pi }}\left( {\frac{{I_{R}} }{{c\rho _{GJ}
\Sigma _{PC} h}}} \right)^{2} \le U\,$$.


{\large{Appendix C. Electromagnetic tornado}}
\medskip

Let us consider an electromagnetic tornado in the gap (see
$^{58}$). The equations of motion in the plane orthogonal to a
magnetic field in complex variables $\xi = x + iy $ take the form
$\frac{{dw}}{{dt}} - i\omega _{c} w = \frac{{eE}}{{mr}}\xi ,\,\,w
= \frac{{d\xi} }{{dt}}$, where $\bf {E}$ $ = E \cdot \frac {\bf
r}{r}, \bf r$$=(x, y)$, is the field of a space charge, and
$\omega _{c} = \frac{{eH}}{{mc}}$ is the cyclotron frequency. In
the axial electric field \textit{E}=\textit{E}(r) depending solely
on the distance \textit{r} from the axis, which we believe to be a
constant parameter, these equations allow solutions of the type
$\xi = \xi _{0} \left( {r} \right)e^{i\,\Omega \,\left( {r}
\right)t}$. The frequencies $\Omega $ obey the equation $\Omega
^{2} - \omega _{c} \,\Omega + \frac{{eE}}{{mr}} = 0$,  roots of
which are equal to $2\Omega _{ \pm} = \omega _{c} \pm \sqrt
{\omega _{c}^{2} - 4\frac{{eE}}{{mr}}} $. Transition to the
l-system in which the bunch moves with relativistic velocity
results in replacing $\Omega \to \Omega /\Gamma $ where $\Gamma $
is the Lorentz-factor. In this derivation, the appearance of the
azimuthal magnetic field in the l-system, which is induced by the
longitudinal electric current should be taken into account. This
field can be found from the Maxwell equations. In the pulsar
conditions there is a small parameter $\frac{{4eE}}{{mr\omega
_{c}^{2}} } < < 1$  the expansion by which yields the values
$\Omega _{ +} \approx \omega _{c}$ and $\Omega _{ -}  \approx
c\frac{{E}}{{H}}\frac{{1}}{{r}}$. The first root is in agreement
with the common cyclotron mode slightly modified by the electric
field. The second root, being of basic interest to us, corresponds
to the drift in the crossed fields. We may check it by going over
to the polar coordinates through substitution $\xi = re^{i\varphi}
$. The equations for the radial and azimuthal velocities $V_{r} =
\frac{{dr}}{{dt}}, V_{\varphi}  = r\frac{{d\varphi} }{{dt}}$ take
the forms $\frac{{dV_{r}} }{{dt}} - \frac{{V_{\varphi} ^{2}}
}{{r}} + \omega _{c} V_{\varphi}  - \frac{{eE}}{{m}} = 0$ and
$\frac{{dV_{\varphi} } }{{dt}} + \frac{{V_{\varphi } V_{r}} }{{r}}
- \omega _{c} V_{r} = 0$, respectively. The stationary solution,
in which the azimuthal velocity is  time-independent, has to
satisfy the conditions $\frac{{V_{\varphi} ^{2}} }{{r}} - \omega
_{c} V_{\varphi}  + \frac{{eE}}{{m}} = 0$ and $\left(
{\frac{{V_{\varphi} }}{{r}} - \omega _{c}}  \right) \cdot V_{r} =
0$. They are held at $V_{r} = 0$ and $V_{\varphi}  = r \cdot
\Omega _{ \pm}  $, where $\Omega _{ \pm}  $ are the frequencies
found above and having the meaning of angular velocities of
rotation (cf. $^{59}$). The solution $V_{\varphi}  = r \cdot
\Omega _{ -} $ describes the electromagnetic tornado in the polar
gap in which the repulsion field of the space charge is
compensated by the Lorentz force, the radial movement is absent,
and rotation is specified by the drift in the crossed fields. Due
to this rotation, the circular polarization appears in the
discharge radiation.

\medskip

{REFERENCES}

\medskip

${1}$. Manchester R.N., Radio Emission Properties of Pulsars. 2.6.
Giant and Not-So-Giant Pulses. In the book "\textit{Neutron Stars
and Pulsars}" W.Becker (Ed.) Springer-Verlag Berlin Heidelberg,
p.33-35 (2009).

${2}$. Kuzmin A.D. Giant pulses of pulsar radio emission.
\textit{Astrophys. \& Space Sci.}, \textbf{308}, 563-567 (2007);
\textit{astro-ph/0701193}

${3}$. Knight H.S. Observational Characteristics of Giant Pulses
and Related Phenomena. \textit{Chin. J. Astron. Astrophys.}
\textbf{6}, Suppl. 2, 41-47 (2006).

${4}$. Soglasnov V.A. Amazing properties of giant pulses and the
nature of pulsar's radio emission. \textit{Proceedings of the 363
WE-Heraeus Seminar on: "Neutron Stars and Pulsars"}, eds. W.
Becker, H.H. Huang, MPE Report 291, pp. 68- 71 (2006);
\textit{astro-ph/0701190}

${5}$. Cairns I.H. Properties and interpretations of giant
micropulses and giant pulses from pulsars. \textit{Astrophys. J.}
\textbf{610}, 948-955 (2004).

${6}$. Staelin D.H. \& Sutton J.M. Observed shape of Crab nebula
radio pulses. \textit{Nature} \textbf{226}, 69-71 (1970).

${7}$. Hankins T.H. \& Eilek J.A. Radio emission signatures in the
crab pulsar. \textit{Astrophys. J.} \textbf{670}, 693 (2007);
\textit{astro-ph/0708.2505}.

${8}$. Hankins T.H., Kern J.S., Weatherall J.C. \& Eilek J.A.
Nanosecond radio bursts from strong plasma turbulence in the Crab
pulsar. \textit{Nature} \textbf{422}, 141-143 (2003).

${9}$. Popov M.V. \& Stappers B. Statistical properties of giant
pulses from the Crab pulsar. \textit{ Astron. \& Astrophys.}
\textbf{470}, 1003-1007 (2007); \textit{astro-ph/ 0704.1197v2}

${10}$. Soglasnov V.A. et al. Giant pulses from PSR B1937+21 with
widths $\leq 15$ nanoseconds and T$_{b}$ $\geq 10^{39}$K, the
highest brightness temperature observed in the Universe.
\textit{Astrophys.J.} \textbf{616}, 439-451 (2004).

${11}$. Kuiper L. et al. Chandra and RXTE studies of the
X-ray/gamma-ray millisecond pulsar PSR J0218+4232.
\textit{Adv.Space Res.} \textbf{33}, 507-512 (2004).

${12}$. Cusumano G. et al. The phase of the radio and X-ray pulses
of PSR B1937+21. \textit{Astron. \& Astrophys.} \textbf{410},
L9-12 (2003); \textit{astro-ph/0309580}

${13}$. Ulyanov O.M., Zakharenko V.V, Deshpande A. et al.
Two-frequency observation of six pulsars using UTR-2 and GEETEE
radio telescopes. \textit {Radio Physics and Radio Astronomy}
\textbf{12}, 5-19 (2007). See also:\\ \textit
{http://www.lorentzcenter.nl/web/2008/306/presentations}.

${14}$. Usov V.V. On two-stream instability in pulsar
magnetosphere. \textit{Astrophys.J.} \textbf{320}, 333-335 (1987).

${15}$. Asseo E., Pelletier G. \& Sol, H. A non-linear radio
pulsar emission mechanism \textit{Mon. Not. R. Astr. Soc.}
\textbf{247}, 529-548 (1990).

${16}$. Weatherall J.C. Modulational instability, mode conversion,
and radio emission in the magnetized pair plasma of pulsars.
\textit{Astrophys. J.} \textbf{483}, 402-413 (1997).

${17}$. Istomin Ya. Origin of Giant Radio Pulses. in \textit{IAU
Symp. 218, Young Neutron Stars \& Their Environments}, ed. F.
Camilo \& B. M. Gaensler (San Francisco: ASP), 62-64 (2003).

${18}$. Lyutikov M. On generation of Crab giant pulses.
\textit{Mon. Not. R. Astr. Soc.} \textbf{381}, 1190 (2007)
\textit{astro-ph/0705.2530}

${19}$. Petrova S.A. On the origin of giant pulses in radio
pulsars., \textit{Astron. \& Astrophys.} \textbf{424}, 224-236
(2004).

${20}$. Ruderman M.A., Sutherland P.G. Theory of pulsars: polar
cap, sparks, and coherent microwave radiation. \textit{Astrophys.
J.} \textbf{196}, 51-72 (1975).

${21}$. Sturrock P.A. A model of pulsars. \textit{Astrophys.J.}
\textbf{164}, 529-556 (1971).

${22}$. Young M.D.T. A Resonant-Mode Model of Pulsar Radio
Emission, in IAU Symp. 218, Young Neutron Stars and Their
Environments\}, eds. F. Camilo \& B.M. Gaensler (San Francisco:
ASP, 2004), p.365; \textit{astro-ph/0310411}

${23}$. Kontorovich V.M. Dice and pulsars\textbf{.}
\textit{Problems of atomic science and technology} No.
3(\ref{eq1}), 195-199 (2007); \textit{astro-ph/0710.4020}

${24}$. Bliokh P.V., Nikolaenko A.P. \& Filippov Yu.F. Global
electro magnetic resonances in the cavity Earth-Ionosphere. Kiev:
Naukova Dumka (1977) 200 p.

${25}$. Kontorovich V.M. \& Flanchik A.B. On the connection
between gamma and radio radiation spectra in pulsars.
\textit{JETP} \textbf{106}, 869-877 (2008);
\textit{astro-ph/0801.0057}

${26}$. Arons J. \& Scharleman E.T. Pair formation above pulsar
polar caps: structure of the low altitude acceleration zone.
\textit{Astrophys.J.} \textbf{231}, 854-879 (1979).

${27}$. Beskin V.S., Gurevich A.G., Istomin Ya.N. Physics of the
pulsar Magnetosphere. Cambridge: Cambridge University Press
(1993).

${28}$. Beskin V.S. MHD Flows in Compact Astrophysical Objects,
Springer (2010) 425 p; Axially Symmetric Stationary Flows in
Astrophysics [in Russian], Moscow: Fizmatlit (2006) 382 p.

${29}$. Kontorovich V.M. On the nature of high brighteness
temperatures of the pulsar giant pulses.
\textit{http://www.ioffe.ru/astro/NS2008/index.html}

${30}$. Muslimov A.G. \& Harding A. K. High-altitude particle
acceleration and radiation in pulsar slot gaps.
\textit{Astrophys.J.} \textbf{606}, 1143-1153 (2004).

${31}$. Ershov A.A. \& Kuzmin A.D. Detection of giant pulses the
pulsar B1112+50. \textit{Astron.Lett.} \textbf{29}, 91-95 (2003).

${32}$. Kondratiev V.I. et al. Detailed study of giant pulses from
the millisecond pulsar B1937+21. Proceedings of the 363.
WE-Heraeus Seminar on: "Neutron Stars and Pulsars", 2006, eds. W.
Becker, H.H. Huang, MPE Report 291, pp. 76-79;
\textit{astro-ph/0701290}

${33}$. Knight H.S., Bailes M., Manchester R.N., Ord S.M. A search
for giant pulses from millisecond pulsars. \textit{Astrophys.J.}
\textbf{ 625}, 951-956 (2005).

${34}$. Eilek J.A. \& Hankins T.H. What makes the Crab pulsar
shine? In: "Forty Years of Pulsars: Millisecond Pulsars, Magnetars
and More", Montreal, August 2007. AIP Conf.Proc. 983, 51;
\textit{astro-ph/0701252}

${35}$. Bilous A.V., Kondratiev V.I., Popov M.V. \& Soglasnov V.A.
Review of overall parameters of giant radio pulses from the Crab
pulsar and B1937+21. In: "Forty Years of Pulsars: Millisecond
Pulsars, Magnetars and More", Montreal, August 2007. AIP
Conf.Proc. 983, 118; \textit{astro-ph/0711.4140}

${36}$. Golitsyn G.S. The place of the Gutenberg-Richter law among
other statistical laws of nature. \textit{Comput. Seismol.}
\textbf{32} 138-162 (2001); Convinience of using energy rather
than mass in the measurement unit system for a number of problems
in physics, mechanics and geophysics. \textit{Phys. Uspekhi}  178,
753 (2008).

${37}$. Romani R.W. \& Jonston S. Giant Pulses from the
millisecond pulsar. \textit{Astrophys.J.Lett.} \textbf{557},
L93-96 (2001).

${38}$. Jonston S. \& Romani R.W. Giant Pulses from PSR B0540-69
in the Large Magellanic Cloud. \textit{Astrophys.J.Lett.}
\textbf{590}, L95-98 (2003).

${39}$. Joshi B.C., Kramer M., Lyne A.G., McLaughlin M. \& Stairs
I.H. Giant Pulses in millisecond pulsars. \textit{Young Neutron
Stars and Their Environments}. \textit{IAU Symp.} \textbf{218},
319-320 (2004).

${40}$. Kuzmin A.D., Ershov A.A. \& Lozovsky B.Ya. Detection of
Giant Pulses from the Pulsar B0031-07. \textit{Astron.Lett.}
\textbf{30}, 247-250 (2004).

${41}$. Ershov A.A. \& Kuzmin A.D.\textbf{} Detection of Giant
Pulses in pulsar PSR J1752+2359.  \textit{A\&A} \textbf{433}, 593
(2005); \textit{astro-ph/0509068}

${42}$. Kuzmin A.D. \& Ershov A.A. Detection of Giant Radio Pulses
from the Pulsar B0656+14. \textit{ Astron.Lett.}  \textbf{32}, 650
(2006); \textit{ astro-ph/0607323}

${43}$. Luo Q. \& Melrose D. Oscillating pulsar polar gaps.
\textit{Mon. Not. R. Astron. Soc}., \textit{ MNRAS } \textbf{387},
1291 (2008); \textit{astro-ph/0804.2009}

${44}$. Popov M., Kuzmin A.D., Ulyanov O.M. et al. Instantaneous
radio spectra of Giant Pulses from the Crab pulsar from decimeter
to decameter wavelengths. \textit{Astron. Rep} \textbf{83},
630-637 (2006); \textit{astro-ph/0606025}

${45}$. Slowikowska A., Jessner A., Kanbach G. \& Klein B.
Comparison of Giant Radio Pulses in Young Pulsars and Millisecond
Pulsars. In: Becker W., Huang H.H. (eds.), Proceedings of the 363
WE-Heraeus Seminar on: Neutron Stars and Pulsars (Posters and
contributed talks) Physikzentrum Bad Honnef, Germany, MPE Report
291, p. 64; \textit{astro-ph/0701105}

${46}$. Jessner A. et al. Giant radio pulses from the Crab pulsar.
\textit{Adv.Space Res.}\textbf{35}, 1166-1171, 2005.

${47}$. Bhat R.N.D., Tingay S.J. \& Knight H.S. Bright giant
pulses from the Crab nebula pulsar: statistical properties, pulse
broadening and scattering due to the nebula. \textit{ ApJ}
\textbf{676}, 1200 (2008); \textit{astro-ph/0801.0334}

${48}$. Popov.M. et al. Multifrequency Study of Giant Radio Pulses
from the Crab Pulsar with the K5 VLBI Recording Terminal.
\textit{astro-ph/0903.2652}

${49}$. Gil J. \&  Melikidze G.I. Angular beaming and giant
subpulses in the Crab pulsar. \textit{Astron. \& Astrophys.}
\textbf{432}, L61-65 (2005).

${50}$. Anderson P.W. Absence of diffusion in certain random
lattices. \textit{Phys. Rev.} \textbf{109}, 1492-1505 (1958).

${51}$. Gredeskul S.A. \& Freilikher V.D. Localization and wave
propagation in randomly-layered media. \textit{Soviet Phys.
Uspekhi} \textbf{160}, 239-262 (1990).

${52}$. Papanicolaou G.C. Wave propagation in a one-dimensional
random medium, SIAM,\textit{ J.Appl.Math.} \textbf{21}, 13-18
(1971).

${53}$. Klyatskin V.I. Stochastic Equations and Waves in
Stochastic Heterogeneous Mediums (in Russian). Moscow: Nauka
(1980). Ondes et equations stochastiques dans les milieux
aleatoirement non homogenes. Les Editions de physique de Besancon
(1985).

${54}$. Lifshits I.M., Gredeskul S.A. \& Pastur L.A. Introduction
to the theory of disordered systems (in Russian). Moscow: Nauka
(1982) 359 p.

${55}$. Volokitin A.S., Krasnosel'skikh V.V. \& Machabeli G.Z.
Waves in the relativistic electron-positron plasma of a pulsar.
\textit{Sov. J. Plasma Phys.} \textbf{11}, 310-338 (1985).

${56}$. Arons J. \& Barnard J.J. Wave propagation in pulsar
magnetospheres: dispersion relations and normal modes of plasmas
in superstrong magnetic fields. \textit{Astrophys.J.}
\textbf{302}, 120-137 (1986).

${57}$. Lominadze J.G., Machabeli G.Z., Melikidze G.I. et al.
Pulsar magnetosphere plasma. \textit{Sov. J. Plasma Phys.}
\textbf{12}, 1233 (1986).

${58}$. Kontorovich V.M. Quantized electromagnetic tornado in
pulsar vacuum gap. \textit{astro-ph/0909.1018}; \textit{JETP}
\textbf{110}, \#6, 966-972 (2010).

${59}$. Davidson R.C. Theory of non neutral plasmas. London:
W.A.Benjamin, Inc. (1974).

\vspace{0,5cm}

\begin{center}
\textbf{Appendix D. Intermediate Epilogue}

\noindent
of a report 2008: arXiv:0911.3272, On high brightness temperature of pulsar giant pulses [A1]. 
\end{center}

\vspace{0,5cm}

Later we have found that in the inner polar gap of pulsar should generate 
strong (coherent) electromagnetic oscillations during acceleration of 
electrons in an increasing (from zero at the surface of the star$^{26}$) 
longitudinal (with respect to the magnetic field) electric $^{A2}$. When 
this field is large enough (narrow internal gap), these oscillations win the 
competition and at sufficiently low frequencies are the dominant mechanism 
of em waves $^{A3}$. Electron acceleration in this case passes through a 
maximum and tends to zero as  the electron velocity approaches to the 
speed of light. Therefore, at high frequencies, that mechanism of radiation 
$^{A2,A3}$ and gives place to other mechanisms of radiation 
(relativistic, beam-plasma, magnetic resonance, etc., see the references in 
the main text and also, for example, in$^{A2}$ ). The emission spectrum when 
averaged over the polar cap becomes a power law with a high-frequency (hf) 
$^{A2}$. This is the physical cause of the power law 
spectrum and its break at radiation due to longitudinal acceleration in the 
gap.

This conclusion is consistent with the Pushchino sample of the Malofeev's 
$^{A4}$ of pulsar spectra with hf break depending on the period as the 
square root of the inverse period of the $^{A5}$. In the theory of 
radiation at longitudinal $^{A2}$ it corresponds to the 
frequency dependence of the hf break as the root of the ratio of 
magnetic field to  period\footnote{ This fact is used by us in$^{A19, A20}$ 
to determine the magnitude of the magnetic field of single pulsars. Assuming 
the magnetic dipole losses thus also possible to determine the angle between 
the magnetic axis and the axis of rotation.}$^)$$^{A2, A3}$. In turn, in the 
theory of the accelerating electric field$^{A6, 28, A7}$ 
such dependence corresponds to a narrow gap, i.e. strong accelerating field. 
The Pushchino sample itself contains enough powerful $^{A8}$. The last 
were specially selected by V.M.Malofeev in compiling the catalog for the 
possibility to trace the features of the spectrum at frequencies far from 
the peak of the radiation, i.e. while their power is substantially reduced. 
The fact that emission of pulsars of the Pushchino sample at low 
frequencies is due to longitudinal acceleration is confirmed also by correlation 
between the hf break and the frequency of maximum of the $^{A5}$, 
which provides a theoretical basis in $^{A3}$. 

On the other hand, for the Crab 
pulsar there is an opportunity to explain the differences in the spectra of 
the main pulse and inter $^{7, 8, 34, A9}$. Really, when changing the mechanisms 
of radiation in the gap with increasing frequency, the relativistic 
mechanism comes into play. And its narrow beam focuses in the direction of the line of sight of inter $^{A10}$ by virtue of the geometry of the pulsar magnetic field. 

It is possible that the emission occurs in narrow $^{28, 
A7}$, as expected in its time in the model of Ruderman and $^{20}$ 
with a strong accelerating field at the surface of the star. However, unlike 
in this model, here the jets appear due to peculiar regime of acceleration 
in the electric field increasing from $^{A3}$: the advantage in the 
acceleration have sections with the higher temperature due to (thermal) 
fluctuations and, consequently, with the greater value of initial electron 
velocity. Current flows warms further this area, and thus must occur 
instability of uniform flow, expressed in the formation of jets. 

Such jets 
are rotating in the crossed fields - magnetic field of the pulsar and electric 
field of the space charge of the same stream, forming an electromagnetic 
"tornado"$^{A1, A11, 58}$. This helps to explain the emergence of circular 
polarization of giant pulses, which can reach 100\% $^{8}$. In favor the 
existence of such "tornado" the appearance of bands in the spectra 
of inter $^{7}$  (for the Crab pulsar) tells us. While in the spectra of the main pulse the bands are not observed. Explanation requires attraction 
(semiclassical) quantization of tornado $^{A12, A13}$. 

Evidence of existence of strong oscillations in the gap can also serve a 
correlation between gamma radiation and giant $^{A14}$. Such a 
correlation component, apparently, is in the gamma-ray emission of the Crab 
$^{A15}$. Gamma radiation in this case arises as a result of the inverse 
Compton scattering of accelerated relativistic electrons at the 
low-frequency oscillations in the $^{25, A16}$. 

The presence of powerful 
low frequency oscillations of the radio band in the inner gap of the pulsar 
is reflected in the processes of particle acceleration and the birth of 
gamma rays in the electric field of the gap, and, consequently, on the birth 
electron-positron plasma in the magnetosphere. The energy density 
oscillations in the gap becomes an important additional parameter on which 
depend the equilibrium conditions in the pulsar. In$^{A17}$ have 
shown that the condition of the birth of electron-positron pairs, taking 
into account the losses on Compton scattering of low-frequency photons in 
the gap, limits on top by the power of low frequency oscillations. Has been 
found the condition that the Compton loss on scattering of low-frequency 
photons are responsible for the shutdown of pulsars. Have been obtained estimates of energy density oscillations in the gap on the plane 
'derivative period - period'. These estimates are consistent with the 
results obtained from the power and spectrum of the observed radio and gamma 
radiation of $^{A17}$. 

Thus appeared a number of additional data, in relation 
to discussed in our $^{A1}$,  that confirm the presence of powerful 
oscillations in the inner ('vacuum') gap. For their existence is not necessarily the 
presence of trapped modes. Other arguments of this $^{A1}$ remain valid. 
New details and discussion, see also in$^{A18}$. 

In the perspective - the 
explanation of the  shift of the phase of inter pulse radiation from its window that is observed on the "middle" radio frequencies in the Crab $^{A9}$. This mysterious effect from our point of view is naturally explained by reflection from the star surface 
the radiation of relativistic particles with a narrow diagram, 
flying across the gap to the surface of the pulsar. It can be as emission of positrons 
 flying to the star from the magnetosphere as well as electrons on 
the "non-preferred" field $^{A7}$. Reflected from the surface of the star 
that radiation is directed at the mirror angle to the magnetic field line on 
which the particles fly. And consequently, to the direction of radiation 
cone of particles flying from the surface and  radiate in the window 
of interpulse. Absence of displacement of interpulse position at lower and 
higher $^{A9}$ is easily attributed to the change of the prevailing 
mechanism of radiation (in this case - the relativistic emission in the 
inner polar gap). Furthermore, it seems quite realistic in this model the 
solution of the problem of coherence emission in relativistic and 
non-relativistic mechanism, which we hope to present on court of readers in 
the nearest future. 

\vspace{0,5cm}

\begin{center}
\textbf{\large{References to Epilogue} }
\end{center}

\vspace{0,5cm}

\noindent
A1. V.M. Kontorovich, On high brightness temperature of pulsar giant pulses; arXiv:0911.3272.

\noindent
A2. V.M. Kontorovich, A.B. Flanchik, High Frequency Cut-off and Changing of Radio Emission mechanism in Pulsars, Astrophys.Space Sci, 345, 169 (2013); astro-ph/1201.0261.

\noindent
A3. V.M. Kontorovich, A.B. Flanchik, Correlation of Pulsar Radio Emission Spectrum with Peculiarities of Particle Acceleration in a Polar Gap. JETP, 2013, Vol. 116, No. 1, pp. 80–86; ЖЭТФ, 143, №1, 92 (2013); arxiv 1210.2858

\noindent
A4. V.M. Malofeev, Catalog radio spectra of pulsars (Pushchino: PRAO ASC FIAN, 1999). In Russian.

\noindent
A5. I. F. Malov, Radio pulsars (M .: Nauka, 2004). In Russian.

\noindent
A6. J. Dyks, B. Rudak, Approximate expressions for polar gap electric field of pulsars 
Astron. \& Astrophys, 362, 1004 (2000); arxiv: astro-ph/0006256.

\noindent
A7. A. K. Harding, The Neutron Star Zoo, Frontiers of Physics, Volume 8, Issue 6, 679, 2013; arxiv: 	astro-ph/1302.0869

\noindent
A8. V.M. Malofeev, in Pulsars: Problems \& Progress, eds S. Jonston, M.A. Walker \& M. Bailes, ASP Confer. Series, 105, 271 (1996).

\noindent
A9. D. Moffett, T. Hankins. Multifrequency Radio Observations of the Crab Pulsar. Astrophys.J. 468, 779 (1996); arXiv:astro-ph/9604163.

\noindent
A10. V.M. Kontorovich, A.B. Flanchik, Physics of neutron stars NS-2011. Book of abstracts, p. 75, http://www.ioffe.ru/astro/NS2011/index.html; JENAM-2011. Book of abstracts, p. 70.

\noindent
A11. V.M. Kontorovich, Giant pulses of pulsars. Problems of atomic science and technology, №4 (68), с.143-148, 2010 In Russian

\noindent
A12. V.M. Kontorovich, Electromagnetic tornado semiclassical quantization and origin of the bandsin the giant pulses frequency spectrum of the Crab pulsar; 	http://www.ioffe.ru/astro/NS2014/ Program (talks and posters).

\noindent
A13. V.M. Kontorovich, The quantization of the electromagnetic tornado and origin of bands 
in the spectrum of giant pulses of the Crab pulsar. PAZH, in press, 2014.

\noindent
A14. A.B. Flanchik, V.M. Kontorovich, Gamma radiation of pulsars as result of inverse compton scattering at acceleration of electrons in a pulsar polar gap, Problems of atomic science and technology, №1 (77), с. 125-129, 2012; http://vant.kipt.kharkov.ua/TABFRAME.html.

\noindent
A15. A.V. Bilous, V.I. Kondratiev, M.A. McLaughlin, et al. Correlation of Fermi photons with highfrequency radio giant pulses from the Crab pulsar // Astrophys. J. 2011, v. 728, p. 110-119.

\noindent
A16. V.M. Kontorovich, A.B. Flanchik, The theory of radio emission in the inner gap and correlation with gamma radiation in pulsars. http://hea.iki.rssi.ru/d/conf/2011/hea/talk/77/

\noindent
A17. V.M. Kontorovich, A.B. Flanchik, The influence of powerful low-frequency oscillations in the vacuum gap on the acceleration of electrons  and the shutdown line in pulsars. (In Russian). Problems of atomic science and technology, №4 (68), с.170-175, 2010. 

\noindent
A18. Jim L. Palfreyman, Aidan W. Hotan, John M. Dickey, Timothy G. Young, Claire E. Hotan, Consecutive Bright Pulses in the Vela Pulsar, Astrophys JL 735, Is. 1, L17, (2011)

\noindent
A19. V.M. Kontorovich, Analogs of the Earth-ionosphere cavity in theories of pulsar radio emission, in Proc. Conf. EMES2012 (Kharkov, 2012)  http://ri.kharkov.ua/emes/EMES2012\_Theses/pdf, 106.

\noindent
A20. V.M. Kontorovich, The magnetic fields of radio pulsars. Astron. Zh. In press, 2014; \\	http://hea.iki.rssi.ru/conf/hea2013/.

\end{document}